\documentclass[11pt,a4paper]{article}

\usepackage{amsmath, amssymb, amsthm}
\usepackage{graphicx}
\usepackage{caption}
\usepackage{subcaption}
\usepackage{geometry}
\usepackage{hyperref}
\usepackage{booktabs}
\usepackage{multirow}
\usepackage{xcolor}
\usepackage{float}
\usepackage{cite}
\usepackage{algorithm2e}
\usepackage{comment}
\usepackage{cancel}
\makeatletter
\renewcommand{\@algocf@capt@plain}{above}
\makeatother
\title{A coupled Kolmogorov-Arnold Network and Level-Set framework for evolving interfaces}
\author{
    Tarus Pande$^{a}$, V\,M\,S\,K Minnikanti$^{a}$, Shyamprasad Karagadde$^{a,b,*}$ \\[0.5em]
    \small $^{a}$ Department of Mechanical Engineering \\
    \small $^{b}$ Centre for Machine Intelligence and Data Science \\
    \small Indian Institute of Technology Bombay, Mumbai 400076, India \\
    \small $^{*}$ Corresponding Author: s.karagadde@iitb.ac.in
}
\date{}

\begin{document}
\maketitle

\begin{abstract}
Kolmogorov–Arnold Networks (KANs) require significantly smaller architectures compared to multilayer perceptron (MLP)-based approaches, while retaining expressive power through spline-based activations. Moving interfaces are ubiquitous in physical systems, whose numerical solutions are quite complex. We propose a shallow KAN framework combined with a Level-set formulation that directly approximates the temperature distribution $T(\textbf{x},t)$ and the moving interface $\Gamma(t)$, enforcing the governing PDEs, phase equilibrium, and Stefan condition through physics-informed residuals. Numerical experiments in one and two dimensions show that the framework achieves accurate reconstructions of both temperature fields and interface dynamics, highlighting the potential of KANs as a compact and efficient alternative for moving boundary PDEs. First, we validate the model with semi-infinite analytical solutions. Subsequently, the model is extended to 2D using a level-set based formulation for interface propagation, which is solved within the KAN framework. This work demonstrates that KANs are capable of solving complex moving boundary problems without the need for measurement data. 
\end{abstract}

\section{Introduction}

Phase-change problems such as melting and solidification are commonly formulated as moving-boundary (Stefan-type) problems, where coupled transport PDEs determine the evolution of an interface separating phases~\cite{dantzig2016solidification}. The interface (boundary dividing different phases) evolution is commonly modelled using a sharp~\cite{SATO2013127, SCAPIN2020109251} or diﬀuse~\cite{PhysRevE.69.051607} or hybrid-approaches~\cite{II20122328}. For Stefan-type problems, stability and accuracy are tightly linked to how well the interface conditions and latent-heat balance are enforced; small temperature errors can induce large interface-location errors over time, motivating alternative space--time formulations beyond conventional discretizations\cite{ASAITHAMBI1992239, GIBOU2002205}.

Physics-Informed Neural Networks (PINNs)~\cite{raissi_physics-informed_2019} approximate the solution with a neural network trained by a loss composed of governing physics and are demonstrated to solve both forward and
inverse problems~\cite{cai_physics-informed_2021, YU2022114823, PATRA2025109854}. Despite their appeal, PINNs face several challenges when extended to solve moving boundary problems. Firstly, they often have to incorporate measurement data to stabilize training, particularly in nonlinear heat transfer and two-phase Stefan problems~\cite{cai_physics-informed_2021}. Secondly, limited by spectral bias, standard PINNs struggle to resolve high-frequency variations~\cite{cao2020understandingspectralbiasdeep, rahaman2019spectralbiasneuralnetworks}, which may affect the accuracy of gradients and sharp-boundary conditions on interfaces. Consequently, solving moving boundary PDEs generally requires deeper network architectures, which increases training cost and hampers the interpretability of the solution~\cite{cuomo_scientific_2022}. In a recent study ~\cite{ANAGNOSTOPOULOS2026107983} investigated the use of PINN for phase transition problems, with
adaptive weighting scheme and residual balancing and require substantial optimization effort and tuning. 

Another study by Mullins et al. \cite{levelsetPINNadv} used MLP-type architecture for level-set equation to handle interfacial dynamics and demonstrated comparable performance similar to conventional techniques but may need  careful tuning of weights and deep networks may demand very high sampling to extend it to coupled Stefan-type interface problems.

In contrast to MLP based architectures, the recently proposed Kolmogorov-Arnold Networks (KANs)~\cite{liu_kan_2024}, grounded in the Kolmogorov-Arnold representation theorem~\cite{kolmogorov1956representation, kolmogorov1957representation}, have better interpretability and accuracy with shallower networks. They incorporate nonlinearity directly into learnable basis functions on the network edges rather than relying on fixed node activations, KANs with shallower networks matched comparable acurracy of standard MLP PINNs~\cite{liu_kan_2024}. Recent studies have demonstrated that KANs effectively mitigate the spectral bias observed in standard MLP-based PINNs, enabling the efficient capture of fine-scale features and sharp gradients~\cite{wang2025expressivenessspectralbiaskans}. In this work, we propose a shallow KAN framework that approximates the temperature distribution and the moving interface by enforcing the governing PDEs, phase equilibrium (dirichlet condition of interface), and Stefan condition through physics-informed residuals. Interface capture is handled using Level Set (LS) method~\cite{osher_sethian_1988} adopted to KAN architecture, termed KANLS methodology.  

To enhance the accuracy near the moving boundary, we employ interface-focused collocation resampling. We validate the framework on benchmark 1D and 2D Stefan problems with available reference solutions (closed-form or high-fidelity numerical), reporting errors in both the temperature field and the inferred interface location, without using measurement data.

\section{Problem Statement \& Methodology}

The Stefan problem is a classical free-boundary problem modeling phase-change
processes such as melting and solidification.
Let $\Omega\subset\mathbb{R}^3$ denote the spatial domain, initially partitioned
into a liquid region $\Omega_\ell(0)$ and a solid region $\Omega_s(0)$, separated
by an interface $\Gamma(0)$. The objective is to determine the temperature fields
$u_\ell(\mathbf{x},t)$ and $u_s(\mathbf{x},t)$ in $\Omega_\ell(t)$ and
$\Omega_s(t)$, respectively, together with the evolution of the moving interface
$\Gamma(t)$.

Heat conduction in each phase is governed by
\begin{align}
\partial_t u_\ell &= \alpha_\ell \nabla^2 u_\ell, 
&& \mathbf{x}\in\Omega_\ell(t), \\
\partial_t u_s &= \alpha_s \nabla^2 u_s, 
&& \mathbf{x}\in\Omega_s(t),
\end{align}
where $\alpha_i = k_i / (\rho_i c_{p,i})$ denotes the thermal diffusivity of phase
$i\in\{\ell,s\}$.

At the interface $\Gamma(t)$, the temperature satisfies the equilibrium condition
\begin{equation}
u_\ell = u_s = T_m,
\qquad \mathbf{x}\in\Gamma(t),
\label{eq:interface_temp}
\end{equation}
and the interface motion is governed by the Stefan condition
\begin{equation}
\rho_s L V_n
=
k_s \nabla u_s \cdot \mathbf{n}
-
k_\ell \nabla u_\ell \cdot \mathbf{n},
\qquad \mathbf{x}\in\Gamma(t),
\label{eq:stefan_balance}
\end{equation}
where $\mathbf{n}$ is the unit normal pointing into the liquid phase and $v_n$ is
the normal interface velocity.

At $t=0$, the temperature field is prescribed as
\begin{equation}
u(\mathbf{x},0) = 
\begin{cases}
u_s^0(\mathbf{x}), & \mathbf{x} \in \Omega_s(0), \\
u_l^0(\mathbf{x}), & \mathbf{x} \in \Omega_l(0),
\end{cases}
\end{equation}
and appropriate boundary conditions are imposed on the external boundary $\partial \Omega$.
\\

In one-dimensional phase-change problems, the solid–liquid interface can be
described explicitly by a single-valued function of time. In contrast, in
multi-dimensional settings the interface may undergo topological changes and
cannot, in general, be represented as a single-valued graph. To handle such
geometrical complexity, we adopt the \textit{level set method}, in which the
moving interface is represented implicitly by a smooth scalar field
$\phi(\mathbf{x},t)$ defined over the spatial domain
$\mathbf{x} \in \mathbb{R}^n$, such that

\begin{equation}
    \Gamma(t) = \left\{ \mathbf{x} \in \mathbb{R}^n : \phi(\mathbf{x},t) = 0 \right\},
\end{equation}
where $\Gamma(t)$ denotes the moving phase boundary. The sign of $\phi$ distinguishes the two regions:
\begin{align}
    \Omega_s(t) &= \{ \mathbf{x} : \phi(\mathbf{x},t) < 0 \}, \\
    \Omega_\ell(t) &= \{ \mathbf{x} : \phi(\mathbf{x},t) > 0 \}.
\end{align} 

In the Stefan problem, this velocity is determined by the balance of heat fluxes across the interface \ref{eq:stefan_balance}. The temperature continuity at the interface is imposed via equation \ref{eq:interface_temp}.

This level set framework provides a unified and dimension-independent representation for the moving interface, enabling straightforward extension to both two- and three-dimensional Stefan problems.

We now generalize the KAN-based solution framework for the $n$-dimensional two-phase Stefan problem.  
The model employs three KAN subnetworks to represent the three unknown fields:$(u_s(\mathbf{x},t), \; u_\ell(\mathbf{x},t), \; \phi(\mathbf{x},t))$,
where $u_s$ and $u_\ell$ are the temperature fields in the solid and liquid phases, and $\phi$ is the level set function defining the moving interface.

The overall loss function $\mathcal{L}$ is composed of multiple residual terms that enforce the PDEs, interface conditions, and physical constraints:
\begin{equation}
    \mathcal{L} = \mathcal{L}_{\text{PDE},s} + \mathcal{L}_{\text{PDE},\ell} + 
    \mathcal{L}_{\text{Interface}} + 
    \mathcal{L}_{\text{Advection}} + 
    \mathcal{L}_{\text{Eikonal}} +
    \mathcal{L}_{\text{BC}} + 
    \mathcal{L}_{\text{IC}}.
\end{equation}

The diffusion equations in the solid and liquid phases are enforced through
smoothly masked PDE residuals,
\begin{align}
\mathcal{L}_{\mathrm{PDE},s}
&=
\left\|
\left(
u_{s,t} - \alpha_s \nabla^2 u_s
\right)
\, H_s(\phi(\mathbf{x},t))
\right\|^2,
\\
\mathcal{L}_{\mathrm{PDE},\ell}
&=
\left\|
\left(
u_{\ell,t} - \alpha_\ell \nabla^2 u_\ell
\right)
\, H_\ell(\phi(\mathbf{x},t))
\right\|^2,
\end{align}
where $H_s$ and $H_\ell$ denote smooth approximations of the characteristic
functions of the solid and liquid regions, respectively. These masks are defined
using a scaled error function,
\begin{subequations}
\label{eq:masks}
\begin{align}
    H_\ell(\phi) &= \frac{1}{2}\left(1 + \text{erf}\left(\frac{\phi}{\epsilon}\right)\right), \label{eq:mask_liquid} \\
    H_s(\phi) &= \frac{1}{2}\left(1 - \text{erf}\left(\frac{\phi}{\epsilon}\right)\right). \label{eq:mask_solid}
\end{align}
\end{subequations}
with $\text{erf}(z) = \frac{2}{\sqrt{\pi}}\int_0^z e^{-t^2} \,dt$ denoting the error function. Since
$\phi(\mathbf{x},t)$ is the signed distance function from the interface,
choosing a small value of $\epsilon$ (here $\epsilon = 0.05$) yields a narrow transition
zone, providing a smooth yet sharp approximation to the indicator functions of
the liquid ($\phi > 0$) and solid ($\phi < 0$) domains.

To weakly enforce temperature continuity across the interface, we define the
interface loss
\begin{equation}
\mathcal{L}_{\mathrm{Interface}}
=
\int_{\Omega}
\left[
\left(u_s(\mathbf{x},t) - T_m\right)^2
+
\left(u_\ell(\mathbf{x},t) - T_m\right)^2
\right]
\, w_\Gamma(\phi(\mathbf{x},t)) \, d\mathbf{x},
\label{eq:interface_loss}
\end{equation}
where the weighting function takes the form of the derivative of the scaled error function defined in Eq.~\eqref{eq:masks}.
\begin{equation}
w_\Gamma(\phi)
\propto \frac{d H}{d \phi}
\implies
w_\Gamma(\phi) = \exp\!\left(-\frac{\phi^2}{\epsilon^2}\right).
\label{eq:interface_weight}
\end{equation}
The Gaussian weighting localizes the loss to a narrow band around the moving
interface $\Gamma(t)=\{\mathbf{x}:\phi(\mathbf{x},t)=0\}$, providing a smooth and
differentiable approximation to the sharp interface temperature 
condition. In order to capture the location of interface, given the interface velocity, the level-set function $\phi(\mathbf{x},t)$ is defined as the signed normal
distance from the solid--liquid interface, with $\phi>0$ in the liquid phase,
$\phi<0$ in the solid phase, and $\phi=0$ on the interface
$\Gamma(t)=\{\mathbf{x}:\phi(\mathbf{x},t)=0\}$. The interface is evolved by
advecting $\phi$ according to
\begin{equation}
\phi_t + F(\mathbf{x},t)\,|\nabla\phi| = 0,
\label{eq:ls_advection}
\end{equation}
where $F$ is a velocity field constructed as an extension of the normal interface
velocity following the classical level-set formulation for Stefan
problems~\cite{PhysRevE.62.2471}.

The normal velocity of the interface is defined only on $\Gamma(t)$ through the
Stefan condition,
\begin{equation}
V_n(\mathbf{x},t)
=
\frac{1}{\rho_s L}
\left(
k_s \nabla u_s \cdot \mathbf{n}
-
k_\ell \nabla u_\ell \cdot \mathbf{n}
\right),
\qquad \mathbf{x}\in\Gamma(t),
\label{eq:stefan_velocity}
\end{equation}
where the unit normal vector is obtained from the level-set function as
\begin{equation}
\mathbf{n}
=
\frac{\nabla\phi}{|\nabla\phi|}.
\label{eq:normal}
\end{equation}

Since the Stefan condition provides the interface velocity only on $\Gamma(t)$,
the velocity field $F$ is constructed by extending $V_n$ off the interface along
normal directions. Because $\phi$ represents the signed distance to the
interface, each collocation point $\mathbf{x}\in\Omega$ is associated with a
unique point on the interface given by the normal projection
\begin{equation}
\mathbf{x}_\Gamma
=
\mathbf{x}
-
\phi(\mathbf{x},t)\,\mathbf{n}(\mathbf{x},t),
\label{eq:normal_projection}
\end{equation}
which satisfies $\mathbf{x}_\Gamma\in\Gamma(t)$. The normal interface velocity is
evaluated at the projected point $\mathbf{x}_\Gamma$ using
\eqref{eq:stefan_velocity}, and the resulting value is assigned to the original
point $\mathbf{x}$, defining the extended velocity field
\begin{equation}
F(\mathbf{x},t)
=
V_n(\mathbf{x}_\Gamma,t).
\label{eq:velocity_extension}
\end{equation}
By construction, $F$ is constant along normal rays to the interface.

The level-set evolution equation~\eqref{eq:ls_advection} is enforced in a weak
sense at all collocation points through the advection loss
\begin{equation}
\mathcal{L}_{\mathrm{Advection}}
=
\left\|
\phi_t
+
F(\mathbf{x},t)\,|\nabla\phi|
\right\|^2.
\label{eq:advection_loss}
\end{equation}

Since the velocity extension relies on $\phi$ remaining a signed-distance
function, the condition $|\nabla\phi|=1$ is weakly enforced via the Eikonal
regularization
\begin{equation}
\mathcal{L}_{\mathrm{Eikonal}}
=
\left\|
|\nabla\phi| - 1
\right\|^2,
\label{eq:eikonal_loss}
\end{equation}
which eliminates the need for explicit reinitialization.

The boundary and initial conditions are imposed weakly through the loss terms
\begin{align}
\mathcal{L}_{\mathrm{BC}}
&=
\mathbb{E}_{\mathbf{x}\in\partial\Omega,\,t}
\left[
\left\| u(\mathbf{x},t) - u_D(\mathbf{x},t) \right\|^2
+
\left\| \nabla u(\mathbf{x},t)\cdot \mathbf{n} - g_N(\mathbf{x},t) \right\|^2
\right],
\\
\mathcal{L}_{\mathrm{IC}}
&=
\mathbb{E}_{\mathbf{x}\in\Omega}
\left[
\left\| u(\mathbf{x},0) - u_0(\mathbf{x}) \right\|^2
+
\left\| \phi(\mathbf{x},0) - \phi_0(\mathbf{x}) \right\|^2
\right],
\end{align}
where $u_D$ and $g_N$ denote prescribed Dirichlet and Neumann boundary data,
respectively, and $u_0$ and $\phi_0$ are the initial temperature and phase
distributions.

Training points are generated as follows: (i) \textbf{Collocation points:} uniformly sampled throughout the spatio-temporal domain, with additional points concentrated near the interface $\Gamma(t)$ to enhance accuracy in regions with steep temperature gradients, (ii) \textbf{Boundary points:} used to impose boundary conditions through $\mathcal{L}_{\text{BC}}$, (iii) \textbf{Initial points:} used to satisfy $\mathcal{L}_{\text{IC}}$, defining the initial temperature distribution and initial interface location.

The training data generation method is summarized below in Algorithm ~\ref{alg:data_gen}.

\vspace{1em}

\begin{algorithm}[H]
\caption{Adaptive Training Data Generation}
\label{alg:data_gen}
\KwIn{Batch size \{$N_{dom}, N_{bc}, N_{ic}$\}, Physics Parameters $\mathcal{P}$, Current Model $\mathcal{M}_{\theta}$}
\KwOut{Training Datasets $\mathcal{D}_{\text{dom}}, \mathcal{D}_{\text{bc}}, \mathcal{D}_{\text{ic}}$}
\BlankLine
Extract domain size $L$ and time range $[t_{\text{start}}, t_{\text{end}}]$ from $\mathcal{P}$\;
\BlankLine
$N_{\text{interface}} \leftarrow 0.3 \times N_{dom}$, $N_{\text{uniform}} \leftarrow N_{dom} - N_{\text{interface}}$\;
Sample set $X_{\text{uniform}}$ of $N_{\text{uniform}}$ points uniformly from $\Omega \times [t_{\text{start}}, t_{\text{end}}]$\;
Sample candidate pool $X_{\text{pool}}$ of $10 \times N_{\text{interface}}$ points uniformly\;
Compute level-set predictions $\hat{\phi} \leftarrow \mathcal{M}_{\phi}(X_{\text{pool}})$\;
Select $X_{\text{interface}} \subset X_{\text{pool}}$ with smallest $|\hat{\phi}|$ such that $\|X_{\text{interface}}\| = N_{interface}$\;
$\mathcal{D}_{\text{dom}} \leftarrow X_{\text{uniform}} \cup X_{\text{interface}}$, enable gradients\;
\BlankLine
Sample $N_{bc}$ points $\mathbf{x}_{\text{bc}}$ on $\partial \Omega$ and compute exact $T_{\text{bc}}$\;
$\mathcal{D}_{\text{bc}} \leftarrow \{(\mathbf{x}_{\text{bc}}, t_{\text{bc}}), T_{\text{bc}}^{\text{exact}}\}$\;
Sample $N_{ic}$ points $\mathbf{x}_{\text{ic}}$ in $\Omega$ at $t_{\text{start}}$ and compute exact $T_0, \phi_0$\;
$\mathcal{D}_{\text{ic}} \leftarrow \{(\mathbf{x}_{\text{ic}}, t_{\text{start}}), T_0, \phi_0\}$\;
\BlankLine
\Return $\{\mathcal{D}_{\text{dom}}, \mathcal{D}_{\text{bc}}, \mathcal{D}_{\text{ic}}$\}
\end{algorithm}
\vspace{2mm}
KANs are motivated by the Kolmogorov–Arnold representation theorem, which states that any continuous multivariate function $f: [0,1]^n \to \mathbb{R}$ can be written in the form
\begin{equation}
    f(x_1, \dots, x_n) = \sum_{q=1}^{2n+1} \Phi_q\!\Bigl( \sum_{p=1}^n \varphi_{q,p}(x_p) \Bigr),
\end{equation}
where each $\varphi_{q,p} : [0,1] \to \mathbb{R}$ and $\Phi_q : \mathbb{R} \to \mathbb{R}$ are univariate continuous functions~\cite{liu_kan_2024}.

In the KAN architecture, this theorem is generalized: instead of fixed linear weights and node‐wise activation functions, \textit{each weight} connecting layer $l$ to $l+1$ is replaced by a learnable univariate function (parametrized by a spline) acting on the input coordinate, and then summed across inputs. Thus, a KAN layer of shape $[m, n]$ can be viewed as a matrix of functions $[\varphi_{ij}]$, where
\begin{equation}
    \Phi(x) = \begin{pmatrix}
    \varphi_{11}(x_1) + \cdots + \varphi_{1m}(x_m) \\
    \vdots \\
    \varphi_{n1}(x_1) + \cdots + \varphi_{nm}(x_m)
\end{pmatrix}.
\end{equation}

KANs stack such layers, allowing compositions of these ``univariate‐edge, sum‐node'' layers to build expressive models. Because the univariate functions (splines) are learned, KANs can adaptively shape nonlinearity per edge, rather than relying on a fixed activation per neuron. Empirically, this leads to more parameter‐efficient representations compared to standard MLPs for tasks including function approximation and PDE solving~\cite{liu_kan_2024}.

The training procedure for the KAN framework is structurally similar in both the one-dimensional and two-dimensional Stefan problems. 
We employ the AdamW optimizer with weight decay regularization and a learning rate scheduler to adaptively reduce the step size when the loss plateaus. 

During optimization, gradient clipping is applied to avoid exploding gradients, and resampling of collocation points is performed iteratively to improve resolution near the moving interface. 
This adaptive resampling ensures that steep gradients at the solid–liquid boundary are accurately captured as training progresses. 
The overall training iteration in the KANLS methodology for the n-dimensional Stefan problem is summarized below in Algorithm ~\ref{alg:train_itr}.

\vspace{1em}

\begin{algorithm}[H]
\caption{Training Iteration for the nD Stefan KAN}
\label{alg:train_itr}
\For{epoch $=1,2,\dots,N$}{
    Sample collocation points $(\mathbf{x_f}, t_f)$, boundary points $(\mathbf{x_b}, t_b)$, and initial points $(\mathbf{x_0}, t_0)$\;
    Compute network predictions $u_1(\mathbf{x}, t), u_2(\mathbf{x}, t), \Gamma(t)$\;
    Evaluate PDE residuals $\mathcal{L}_{\text{PDE},l}, \mathcal{L}_{\text{PDE},s}$, interface loss $\mathcal{L}_{\text{Stefan}}$, \\
    continuity losses $\mathcal{L}_{\text{Cont},u_1}, \mathcal{L}_{\text{Cont},u_2}$, boundary loss $\mathcal{L}_{\text{BC}}$, and initial loss $\mathcal{L}_{\text{IC}}$\;
    Form total loss $\mathcal{L} = \sum_i w_i \mathcal{L}_i$\;
    Compute gradients $\nabla_{\boldsymbol{\theta}} \mathcal{L}$ and apply gradient clipping\;
    Update parameters $\boldsymbol{\theta} \leftarrow \boldsymbol{\theta} - \eta \nabla_{\boldsymbol{\theta}} \mathcal{L}$ (AdamW)\;
}
\end{algorithm}

\section{Results}

\subsection{1D Stefan Problem: Temperature and Interface Evolution}

We first validate the proposed KAN framework on the one-dimensional two-phase
Stefan problem. The spatial domain is the unit interval
$\Omega = [0,1]$, initially divided into solid and liquid regions separated by an
interface located at $x = s_0$.

The governing conditions are: Initial condition (IC): $u(x,0) = T_1$ for $x < s_0$ and $T_2$ for $x \geq s_0$, where $T_1 < T_m < T_2$; Boundary conditions (BC): $u(0,t) = T_1$, $u(1,t) = T_{\text{exact}}(1,t)$, where $T_{\text{exact}}(x,t)$ denotes the analytical solution of the semi-infinite Stefan problem evaluated at the right boundary.

For this problem, three separate KAN subnetworks are employed: the solid-phase ($u_s(x, t)$) and liquid-phase ($u_\ell(x, t)$) temperature networks, both with a $[2, 4, 4, 1]$ architecture, and the interface position network $s(t)$, which has a $[1, 2, 2, 1]$ architecture.

This configuration was selected based on empirical evaluation and was found to
provide adequate representational capacity for capturing the temperature
evolution and interface motion in one dimension.

\begin{figure}[H]
    \centering
    \begin{subfigure}{0.48\textwidth}
        \centering
        \includegraphics[width=\linewidth]{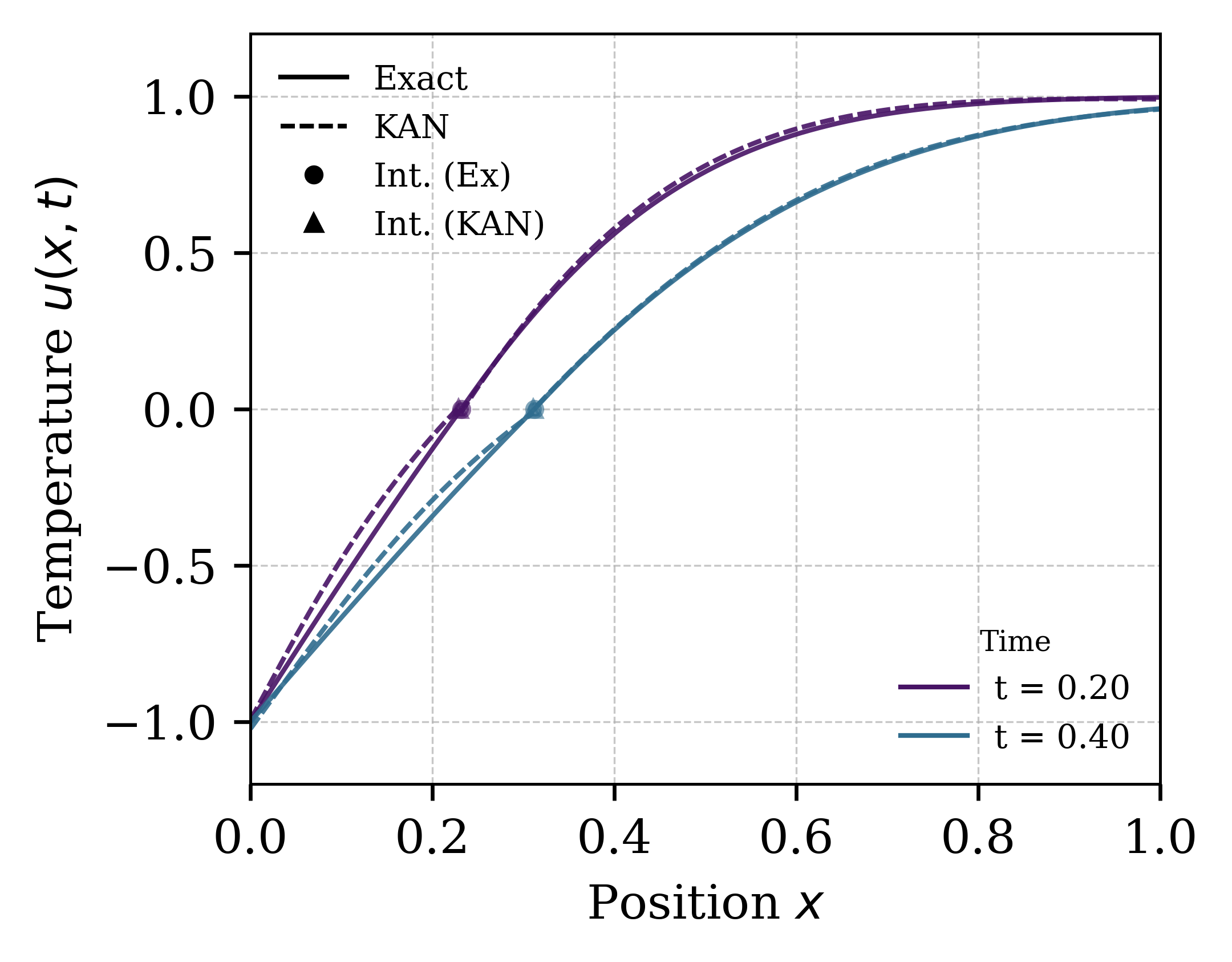}
        \caption{Temperature Evolution}
        \label{fig:1d_temp_evolution}
    \end{subfigure}
    \hfill
    \begin{subfigure}{0.48\textwidth}
        \centering
        \includegraphics[width=\linewidth]{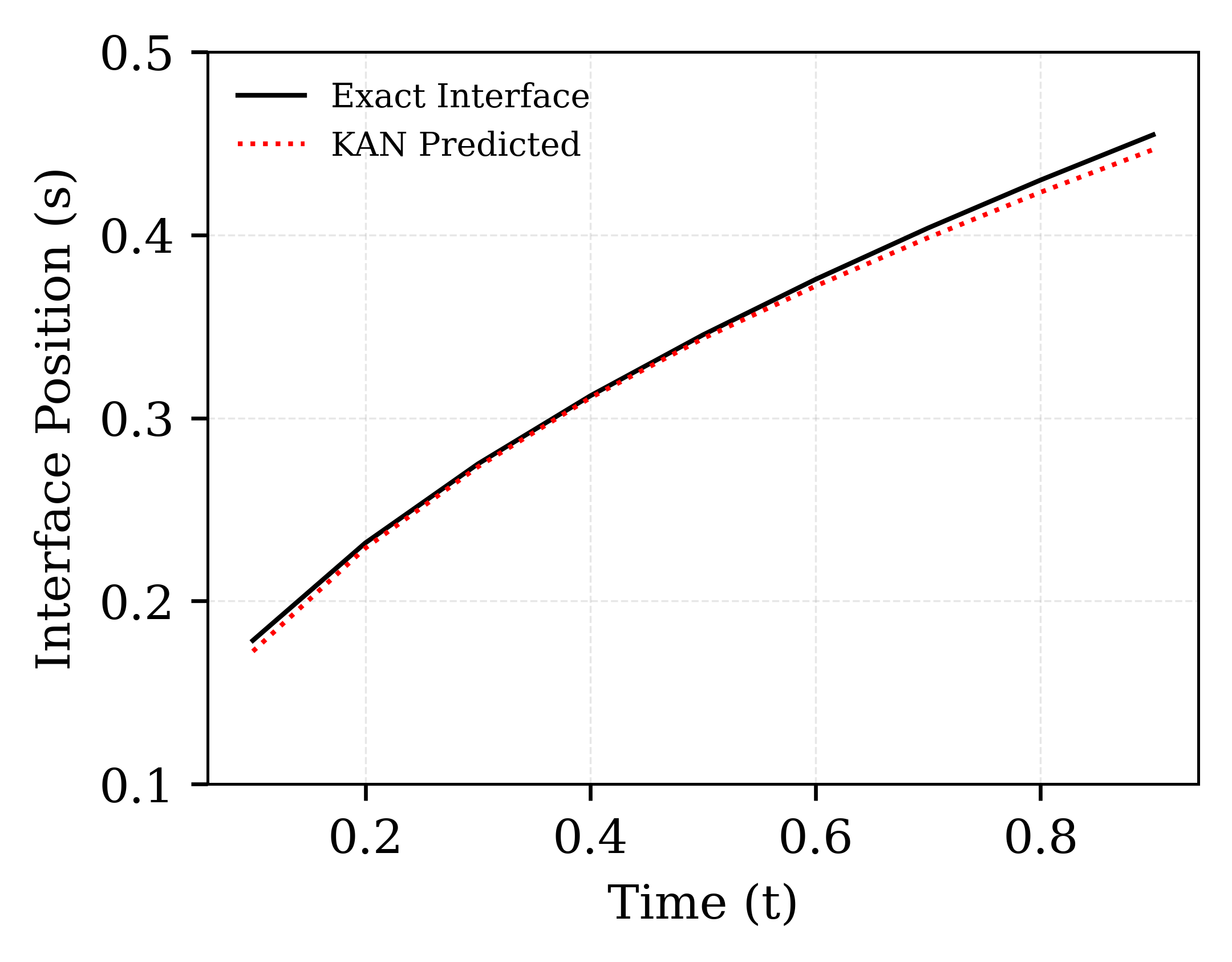}
        \caption{Interface Evolution}
        \label{fig:1d_interface_evolution}
    \end{subfigure}
    \caption{KAN vs. exact solution: (a) temperature, (b) interface.}
    \label{fig:1d_error_comparison}
\end{figure}

Figure~\ref{fig:1d_error_comparison} shows the predicted temperature profiles in both phases at different time instances, along with the evolution of the moving interface compared to the analytical solution. 
The results demonstrate that the network accurately captures both the smooth temperature fields and the sharp interface dynamics, confirming the effectiveness of the proposed KAN-based formulation for the 1D Stefan problem.

\begin{table}[H]
    \centering
    \caption{Error metrics for temperature and interface predictions at different numbers of collocation points.}
    \label{tab:collocation_errors}
    \renewcommand{\arraystretch}{1.2}
    \setlength{\tabcolsep}{5pt}
    \begin{tabular}{c|cccc|cccc}
        \hline
        \multirow{2}{*}{$n_{\text{coll}}$} & 
        \multicolumn{4}{c|}{\textbf{Temperature}} & 
        \multicolumn{4}{c}{\textbf{Interface}} \\
        \cline{2-9}
         & MAE & MSE & RMS & $R^2$ & MAE & MSE & RMS & $R^2$ \\
        \hline
        500  & 0.0432 & 0.00310 & 0.0556 & 0.9913 & 0.0183 & 0.00034 & 0.0183 & 0.9670 \\
        1000 & 0.0278 & 0.00170 & 0.0413 & 0.9949 & 0.0084 & 0.00009 & 0.0093 & 0.9878 \\
        2000 & 0.0272 & 0.00153 & 0.0391 & 0.9957 & 0.0067 & 0.00005 & 0.0072 & 0.9949 \\
        4000 & 0.0229 & 0.00122 & 0.0349 & 0.9966 & 0.0045 & 0.00003 & 0.0054 & 0.9971 \\
        \hline
    \end{tabular}
\end{table}

For a problem of this nature, the architecture typically utilizes a five-layer MLP with
approximately 100 neurons per layer for each  of the subnetworks $(u_s, u_\ell, s)$, resulting in roughly $1.2\times 10^5$ trainable parameters~\cite{kathane2024physicsinformedneuralnetwork}, whereas the corresponding KAN-based formulation uses only $640$ parameters.

\subsection{2D Stefan Problem: Temperature and Interface Evolution}

We now consider a two-dimensional Stefan problem with a circular (Frank-type)
interface in order to assess the convergence and robustness of the proposed KAN
framework under curved phase boundaries.

The computational domain is defined as $\Omega = [-5,5]\times[-5,5]$, containing an initially circular solid region centered at the origin with radius $R_0 = 1.56$. This setup corresponds to a radially symmetric Stefan problem embedded in a sufficiently large finite domain to approximate far-field conditions.

Initially, the system consists of an inner solid region ($\|\textbf{x}\| \leq R_0$) at the melting temperature $T_m$ and an exterior subcooled liquid at $T_\infty$. Boundary conditions are imposed such that the temperature approaches $T_m$ in the far field, enforced on $\partial\Omega$ via the analytical Stefan solution, while the symmetry at the origin is satisfied by fixing $u(0, t) = T_m$.

Three separate KAN subnetworks were employed for this problem as well: 
the solid-phase ($u_s(\textbf{x}, t)$) and liquid-phase ($u_\ell(\textbf{x}, t)$) temperature networks, both with a $\textbf{[3, 8, 8, 1]}$ architecture, and the
level set network $\phi (\textbf{x}, t)$, which also has a $\textbf{[3, 8, 8, 1]}$ architecture.

\begin{figure}[H]
    \centering
    \begin{subfigure}[t]{0.32\textwidth}
        \centering
        \includegraphics[width=\linewidth]{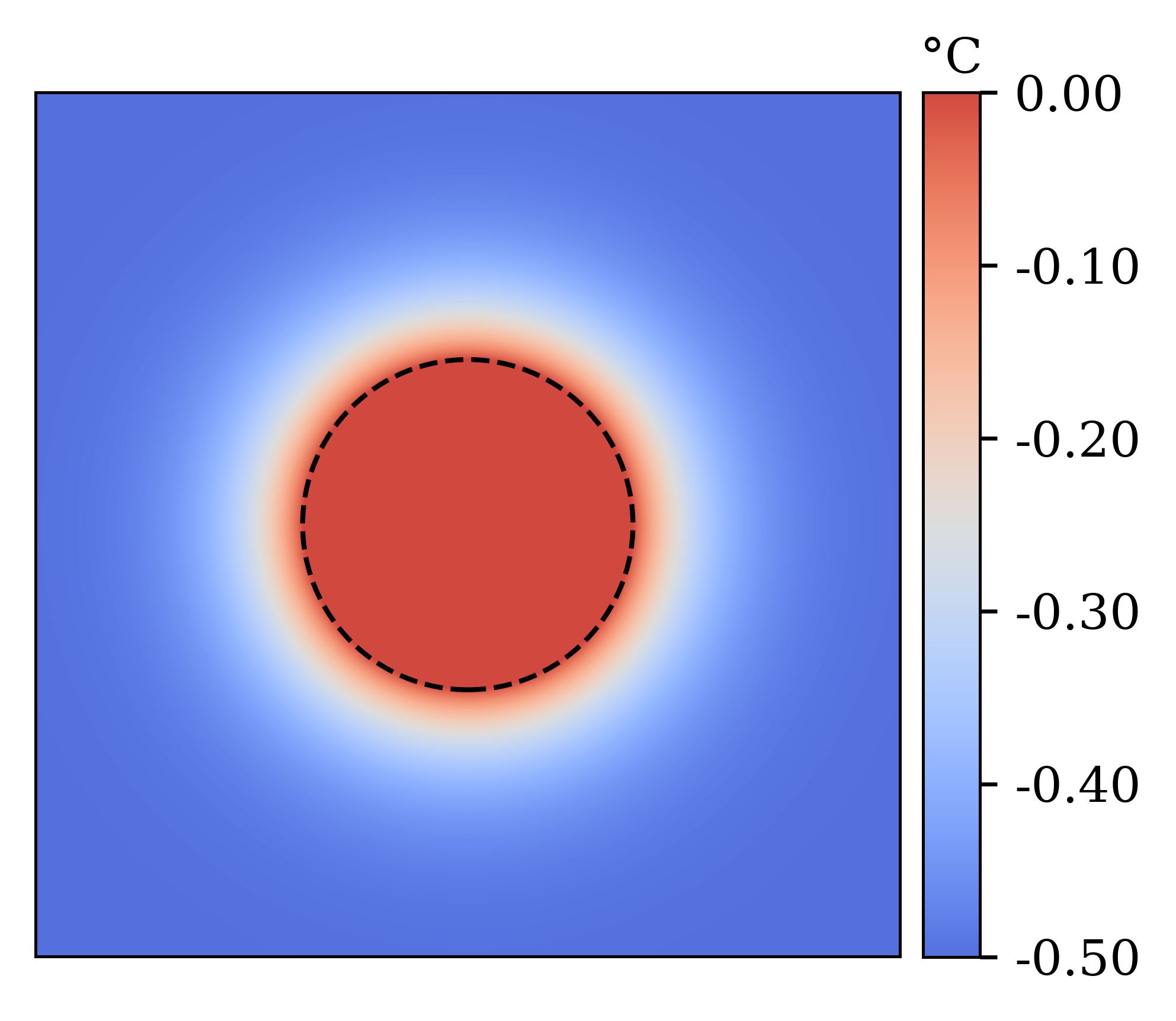}
        \caption{Exact solution}
        \label{fig:exact_solution}
    \end{subfigure}
    \hfill
    \begin{subfigure}[t]{0.32\textwidth}
        \centering
        \includegraphics[width=\linewidth]{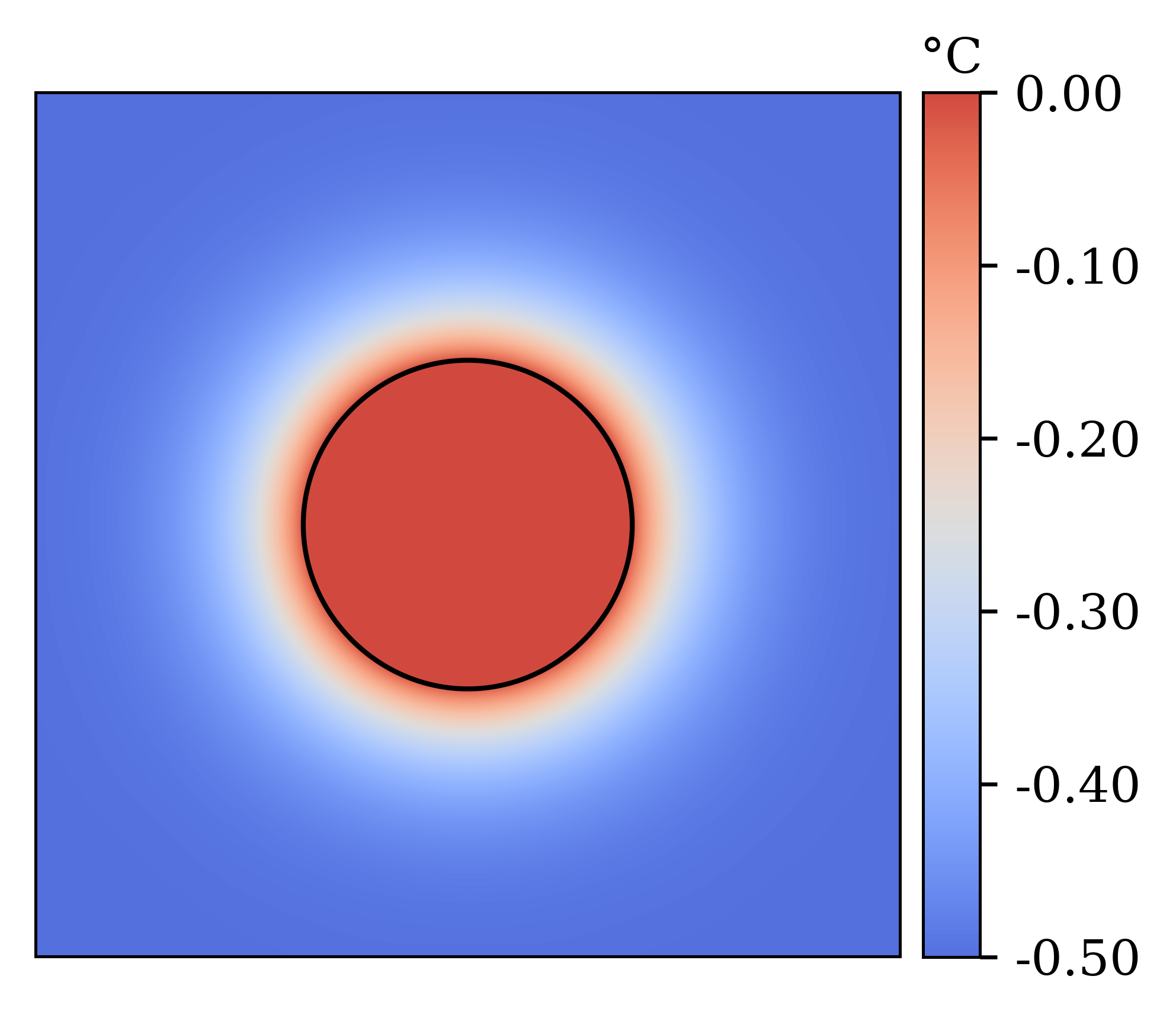}
        \caption{KAN prediction}
        \label{fig:kan_prediction}
    \end{subfigure}
    \hfill
    \begin{subfigure}[t]{0.32\textwidth}
        \centering
        \includegraphics[width=\linewidth]{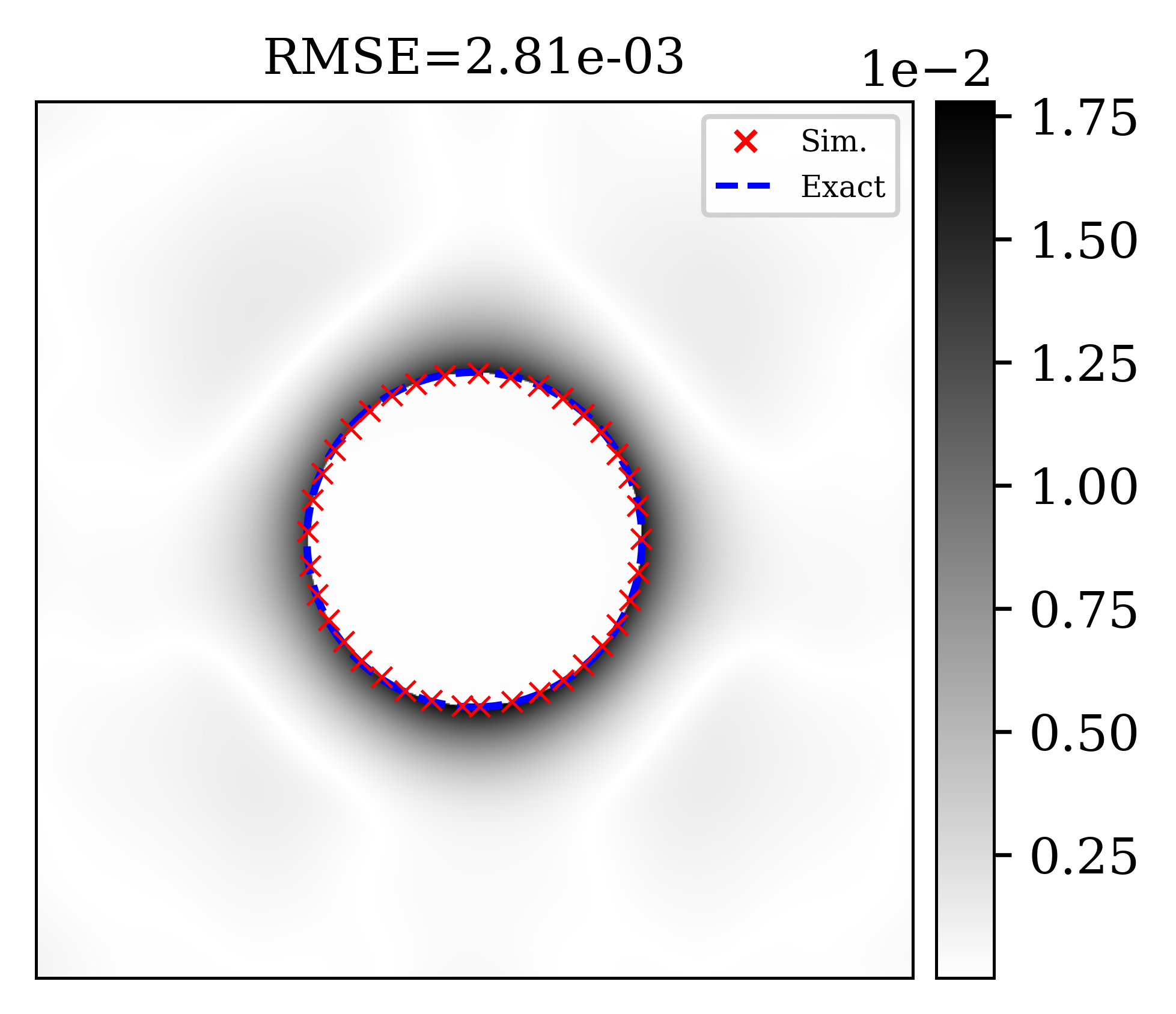}
        \caption{Error map}
        \label{fig:error_map}
    \end{subfigure}
    \caption{Evolution of the temperature field and error distribution for $n_{\text{coll}} = 8000$ at $t = 1.5$.}
    \label{fig:field_evolution}
\end{figure}

Figure~\ref{fig:field_evolution} shows the predicted temperature fields in both phases at $t=1.5$ compared to the analytical solution. 

\begin{figure}[H]
    \centering
    \begin{subfigure}[b]{0.49\textwidth}
        \centering
        \includegraphics[width=\textwidth]{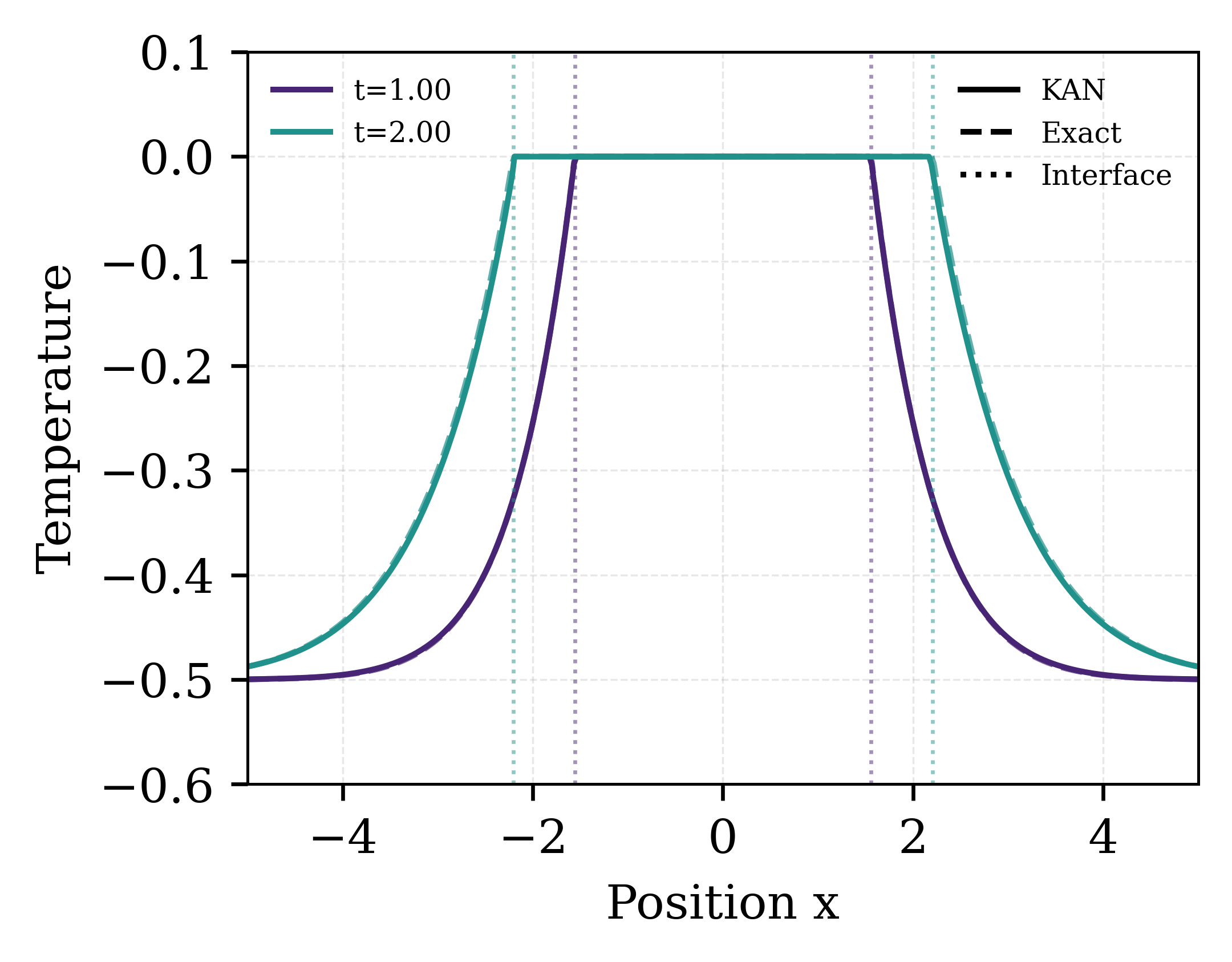}
        \caption{Temperature profiles along $y=0$}
        \label{fig:profiles}
    \end{subfigure}
    \hfill
    \begin{subfigure}[b]{0.49\textwidth}
        \centering
        \includegraphics[width=\textwidth]{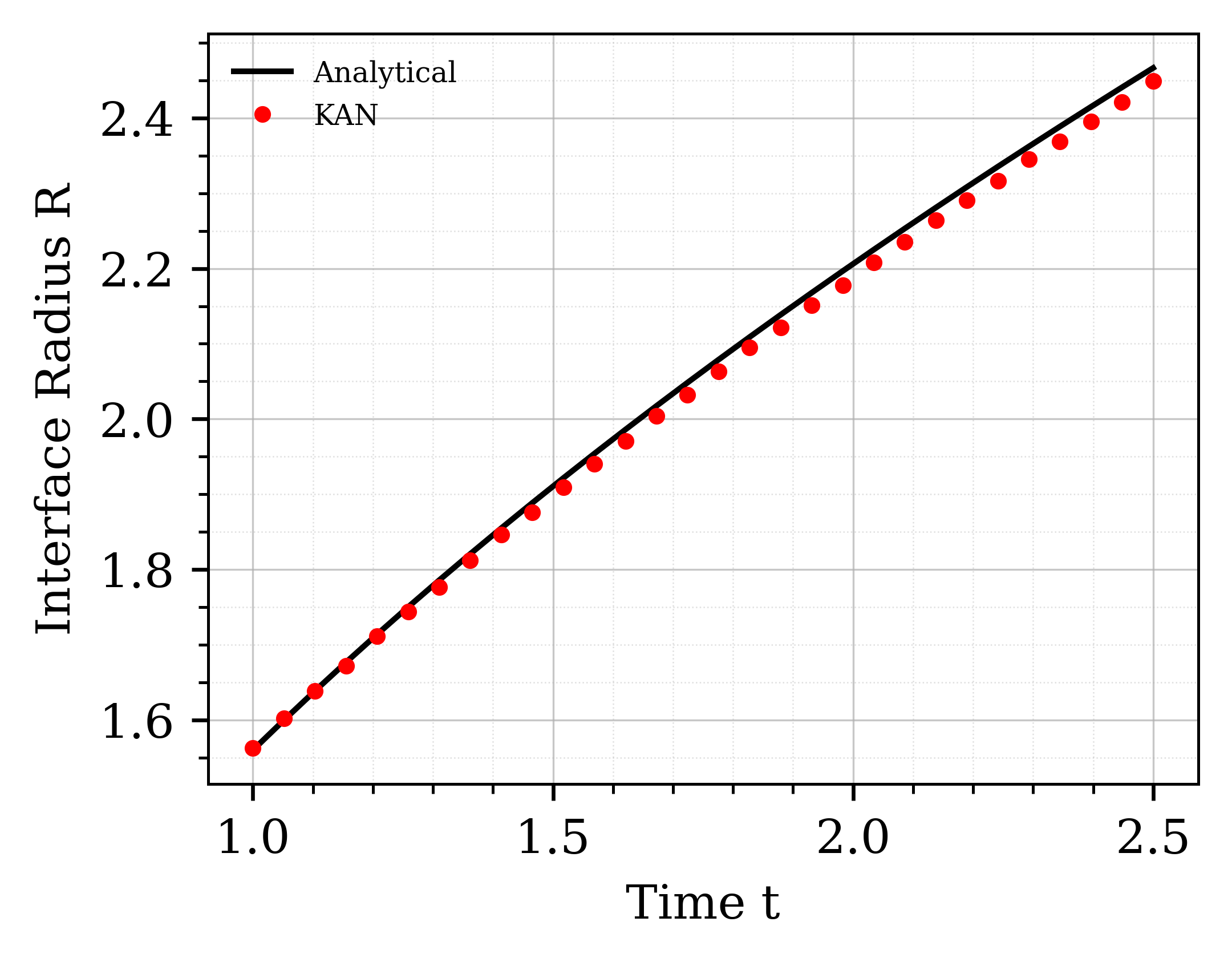}
        \caption{Interface growth dynamics}
        \label{fig:interface_growth}
    \end{subfigure}
    \caption{KAN vs. exact solution: (a) temperature, (b) interface.}
    \label{fig:2d_error_comparison}
\end{figure}

The quantitative accuracy of the KAN framework is further analyzed in Figure~\ref{fig:2d_error_comparison}. The temperature profiles along the symmetry axis $y=0$, shown in Figure~\ref{fig:profiles}, align closely with the exact solution at $t=1.0$ and $t=2.0$. The model correctly captures the expanding isothermal solid core ($u=0$) and the smooth thermal decay in the liquid region, resolving the sharp phase boundary without significant oscillations. Furthermore, the interface dynamics presented in Figure~\ref{fig:interface_growth} verify that the predicted radius $R(t)$ tracks the analytical interface $R(t) = R_0 \sqrt{t}$ with high precision. This confirms that the physics-informed loss functions successfully enforce the Stefan condition governing the interface velocity.

We investigate the impact of collocation density by comparing models trained with varying numbers of collocation points, using space–time pointwise error metrics (MAE, MSE, and $R^2$) computed against the analytical solution (signed distance function for $\phi$).

\begin{table}[H]
    \centering
    \caption{Error metrics for temperature and level set predictions for models trained with different numbers of collocation points.}
    \label{tab:collocation_errors_2d}
    \renewcommand{\arraystretch}{1.2}
    \setlength{\tabcolsep}{6pt}
    \begin{tabular}{c|ccc|ccc}
        \hline
        \multirow{2}{*}{$n_{\text{coll}}$} & 
        \multicolumn{3}{c|}{\textbf{Temperature}} & 
        \multicolumn{3}{c}{\textbf{Level Set}} \\
        \cline{2-7}
         & MAE & MSE & $R^2$ & MAE & MSE & $R^2$ \\
        \hline
        500  & $3.74\times10^{-3}$ & $5.89\times10^{-5}$ & 0.9982 & $4.18\times10^{-2}$ & $2.05\times10^{-3}$ & 0.9990 \\
        1000 & $2.99\times10^{-3}$ & $3.68\times10^{-5}$ & 0.9988 & $3.18\times10^{-2}$ & $1.22\times10^{-3}$ & 0.9994 \\
        2000 & $2.76\times10^{-3}$ & $3.13\times10^{-5}$ & 0.9990 & $2.99\times10^{-2}$ & $1.05\times10^{-3}$ & 0.9995 \\
        4000 & $2.13\times10^{-3}$ & $1.71\times10^{-5}$ & 0.9995 & $2.19\times10^{-2}$ & $5.46\times10^{-4}$ & 0.9997 \\
        8000 & $1.47\times10^{-3}$ & $8.98\times10^{-6}$ & 0.9997 & $1.54\times10^{-2}$ & $2.80\times10^{-4}$ & 0.9999 \\
        16000 & $1.39\times10^{-3}$ & $8.43\times10^{-6}$ & 0.9997 & $1.49\times10^{-2}$ & $2.79\times10^{-4}$ & 0.9999 \\
        \hline
    \end{tabular}
\end{table}

We observe diminishing returns beyond $n_{coll} = 8000$, indicating that the learned solution has effectively converged with respect to collocation density.

\section{Conclusion}
A key motivation for employing Kolmogorov–Arnold Networks (KANs) in the present work is their ability to represent nonlinear mappings with significantly fewer trainable parameters than conventional multilayer perceptrons (MLPs).
KANs construct multivariate functions through compositions of adaptive univariate functions, which has been shown to improve parameter efficiency and reduce redundancy compared to standard fully connected architectures. While KANs have been used for solving PDEs, the present work specifically demonstrates their relevance and superiority in handling moving boundary problems, with interface jump conditions. This overcomes the existing limitation on employing measurement data. Further, we show that the shallow networks with two hidden layers and tens of learnable parameters are sufficient, even without using measurement data, in contrast to physics-informed MLP-based formulations involving nearly a million trainable parameters. The KANLS methodology has a significant potential towards initiating further research in this field involving multi-phase physical systems.  

\bibliographystyle{unsrt}
\bibliography{references}

\end{document}